\documentclass[12pt,aps,showpacs]{revtex4}
\usepackage{amssymb,amsfonts,amsmath}
\makeatletter
\renewcommand{\@makefntext}[1]{\parindent=1em\noindent\hbox to 1.8em{\hss$^{\@thefnmark}$}#1}
\renewcommand{\@footnotemark}{\hbox{\mathsurround=0pt$^{\@thefnmark}$}}

\makeatother
\newcommand{\be}{\begin{equation}}
\newcommand{\bea}{\begin{eqnarray}}
\newcommand{\ee}{\end{equation}}
\newcommand{\eea}{\end{eqnarray}}
\newcommand{\ds}{\displaystyle}
\newcommand{\vph}{\ds\vphantom{\ds\frac{\int_0^1}{\int_0^1}}}

\newcommand{\bp}{\begin{picture}}
\newcommand{\ep}{\end{picture}}
\begin{document}

\title{Spectra and decays of hybrid charmonia}

\author{Yu. S. Kalashnikova}
\affiliation{Institute of Theoretical and Experimental Physics, 117218, B.Cheremushkinskaya 25, Moscow, Russia}

\author{A. V. Nefediev}
\affiliation{Institute of Theoretical and Experimental Physics, 117218, B.Cheremushkinskaya 25, Moscow, Russia}

\begin{abstract}
QCD string model is
employed to calculate the masses and spin splittings of lowest charmonium
hybrid states with a magnetic gluon. Relative decay rates into
various $S$-- and $P$--wave D--meson pairs are calculated for these hybrids.
\end{abstract}

\maketitle

\section{Introduction}

There exist strong arguments in favour of hybrid assignment
for the recently observed $Y(4260)$ state \cite{Yobservation}. Indeed,
this $Y$--meson is definitely a vector one, as it is seen in the
initial state radiation process
$e^+e^- \to \gamma \pi^+\pi^-J/\psi$, but its $e^+e^-$ width
is too small for a conventional $c\bar{c}$ vector, and there is no
visible decay into $D \bar D$ pairs, in spite of the large phase space
available. It is the latter feature that has prompted the hybrid
interpretation of the $Y(4260)$\cite{hybrint}, as the selection rule is established --- see, for example,
\cite{orsay,orsay2,myhybrdecay,kou,hybrdecayflux} --- which forbids the decay
of the vector hybrid into a $D^{(*)} \bar D^{(*)}$ final state.

Competing models for the $Y(4260)$ exist. One is the
$[cs]-[\bar c \bar s]$ diquark--antidiquark model \cite{Maiani}. On the
other hand, the $Y(4260)$ is not far from the $D \bar D_1$
threshold, where $D_1$ is a $P$--wave $1^{+}$ charmed meson, so the $Y(4260)$ state
could be associated with the opening
of a new $S$--wave  $D \bar D_1$ threshold \cite{Rosner}. In this regard
it is important to assess the consequences of the hybrid assignment for
the $Y$.

Hybrids can be considered as bound states of a quark--antiquark pair and
a gluon with quantum numbers
\be
P=(-1)^{l_{q\bar{q}}+j},\quad C=(-1)^{l_{q\bar{q}}+s_{q\bar{q}}+1},
\label{JPCmag}
\ee
for the magnetic gluon ($l_g=j$), and
\be
P=(-1)^{l_{q\bar{q}}+j+1},\quad C=(-1)^{l_{q\bar{q}}+s_{q\bar{q}}+1},\\
\label{JPCel}
\ee
for the electric gluon ($l_g=j \pm 1$),
where $l_g$ is the relative angular momentum between the $q\bar{q}$ pair and
the gluon, $j$ is the total angular momentum of the gluon, $l_{q\bar{q}}$ is the
orbital momentum in the quark--antiquark subsystem, and $s_{q\bar{q}}$ is the spin of the quark--antiquark pair. For a magnetic gluon, the lowest states
correspond to $l_{q\bar{q}}=0$, with the $1^{--}$ hybrid being a
spin--singlet state with respect to the quark spin, while the $J^{-+}$, $J=0,1,2$,
hybrids being spin triplets. These four states are expected to be degenerate in the heavy--quark limit,
with the degeneracy removed by spin--dependent quark--gluon interactions.
In other words, if $Y(4260)$ is indeed a $c\bar{c}g$ vector hybrid, three $J^{-+}$
hybrid charmonia, including the exotic $1^{-+}$ one, should reside not
very far from it. Decays of all these states obey the
above--mentioned selection rule.

It is the latter feature, which makes these states potentially interesting.
Indeed, the same quantum numbers can be also achieved with the electric gluon, $1^{--}$ hybrid
being a spin--triplet state with respect to the quark spin, and $J^{-+}$ hybrids being spin--singlets.
However, hybrids with the electric gluon couple too strongly to two $S$--wave final--state mesons
($D^{(*)} \bar D^{(*)}$) and,
as estimated in Ref.~\cite{orsay2}, do not exist as resonances. On the contrary, for hybrids with the magnetic gluon,
the $D^{(*)} \bar D^{(*)}$ modes are forbidden, and the lowest possible open--charm modes are
the ones with an $S$--wave $D^{(*)}$--meson and a $P$--wave $D_J$--meson, with the thresholds being close
to the masses of such hybrids. Due to a limited phase space, one expects a considerable suppression
of the corresponding decay width, so that hybrids with the magnetic gluon could manifest themselves as resonant states,
and, as such, could be of immediate relevance to the charmonium spectroscopy issues.

In the present paper we calculate spin splittings in the $c\bar{c}g$
system with a magnetic gluon in the framework of the QCD string model
based on the Field Correlator Method (FCM) for QCD (for a
review of the FCM see Ref.~\cite{Simonov}). In this
method, confining dynamics is encoded in gluonic field correlators which
are responsible for area law asymptotic for the Wilson loop.
Starting from the Feynman--Schwinger representation for the quark and gluon
propagators in the external field, one can extract hadronic Green's
functions and calculate the spectra.
The QCD string model corresponds to the limit of a small (vanishing)
gluonic correlation length --- the so-called string limit of QCD.
Then the effective string--type Lagrangian of a colourless object (meson, baryon, hybrid, and so on) can be derived.
The QCD string model was successfully
applied to calculate spectra and other properties of $q\bar{q}$ mesons,
see, for example, Refs.~\cite{KNS,alla}. To account for hybrid excitations one
populates the QCD string with constituent perturbative gluons
\cite{hybrids1}. This approach was used before in order to
investigate the properties of hybrids
\cite{hybrids2}, and the form of the static interquark potentials
in hybrids was studied in detail and compared with the lattice
simulations in Ref.~\cite{hybrids3}.

\section{The Model}

\unitlength=0.5mm
\begin{figure}[t]
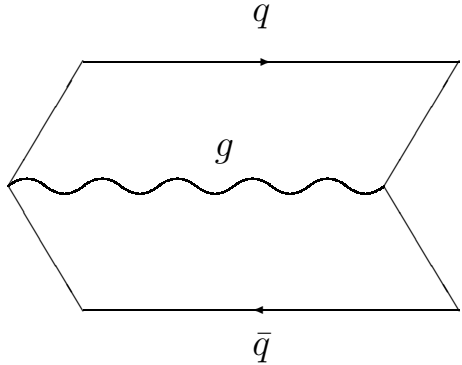

\bp(0,80)
\put(-60,40){\bp(71,51)%
\multiput(0,0)(20,0){5}{
\qbezier(0,0)(5,4)(10,0)
\qbezier(10,0)(15,-4)(20,0)
}
\put(0,0){\line(3,5){20}}
\put(100,0){\line(3,5){20}}
\put(20,33){\line(1,0){100}}
\put(0,0){\line(3,-5){20}}
\put(100,0){\line(3,-5){20}}
\put(20,-33){\line(1,0){100}}
\put(60,33){\vector(1,0){10}}
\put(75,-33){\vector(-1,0){10}}
\put(65,44){\large $q$}
\put(65,-45){\large $\bar{q}$}
\put(55,8){\large $g$}
\ep
}
\ep
\caption{Wilson loop configuration corresponding to the propagation of the
hybrid state.}\label{wl}
\end{figure}

In the framework of the FCM, hybrid is viewed as a gluon with two
fundamental strings attached, with the quark and the antiquark at the ends (see Fig.~\ref{wl}).
Thus the starting point of our analysis is the effective Lagrangian of such a system:
\begin{eqnarray}
L=-\frac{m_q^2}{2\mu_q}-\frac{m_{\bar{q}}^2}{2\mu_{\bar{q}}}&-&\frac{\mu_q+\mu_{\bar{q}}+\mu_g}{2}
+\frac{\mu_q \dot{r}_q^2 + \mu_{\bar{q}}\dot{r}_{\bar{q}}^2+\mu_g\dot{r}_g^2}{2}\nonumber\\
&-&\sigma |\vec{r}_q-\vec{r}_g|\int^1_0 d\beta_1\sqrt{1-l_1^2}-\sigma |\vec{r}_{\bar{q}}-\vec{r}_g|\int^1_0 d\beta_2\sqrt{1-l_2^2},
\label{lagrangian}\\
\vec{l}_1&=&\frac{\vec{r}_q-\vec{r}_g}{|\vec{r}_q-\vec{r}_g|}\times(\beta_1\dot{\vec{r}}_q+(1-\beta_1)\dot{\vec{r}}_g),\nonumber\\
\vec{l}_2&=&\frac{\vec{r}_{\bar{q}}-\vec{r}_g}{|\vec{r}_{\bar{q}}-\vec{r}_g|}\times(\beta_2\dot{\vec{r}}_{\bar{q}}+
(1-\beta_2)\dot{\vec{r}}_g),\nonumber
\end{eqnarray}
where $\vec {r}_q$, $\vec {r}_{\bar q}$, and $\vec {r}_g$ are the quark,
the antiquark, and the gluon coordinates, $m_q$ and $m_{\bar q}$ are the quark and the
antiquark current masses (the gluon is massless), dots stand for time derivatives, and
the minimal surface is approximated by the straight--line string.
The Lagrangian
(\ref{lagrangian}) is written in the einbein field form \cite{einbein};
the einbein fields (or simply einbeins) $\mu_q$, $\mu_{\bar q}$, and $\mu_g$ are introduced to deal with
relativistic kinematics. No time derivatives of the einbeins enter the
Lagrangian, and the corresponding equations of motion yield second--class
constraints (see Ref.~\cite{Dirac} for the details of the constrained systems formalism and for the corresponding terminology):
\be
\frac{\partial L}{\partial \mu_q}=\frac{\partial
L}{\partial \mu_{\bar q}}=\frac{\partial L}{\partial
\mu_g}=0.
\label{lconstraints}
\ee
In principle, one can eliminate einbeins using Eq.~(\ref{lconstraints}).
However, only the einbein form of the Lagrangian provides a meaningful dynamics for a
massless gluon. Application of the einbein field formalism
to the QCD string was suggested in Ref.~\cite{DKS}, and further developments
can be found in Ref.~\cite{KNS}.

Starting from the Lagrangian (\ref{lagrangian}) one can arrive at the
Hamiltonian of the $q\bar{q} g$ system. The general procedure outlined in Refs.~\cite{DKS,KNS} is rather complicated
due to the presence of square roots in (\ref{lagrangian}). However, if one is
interested in the low--lying part of the spectrum, the potential--type
regime can be considered. To this end notice that the
angular velocities $l_{1,2}^2$ in the Lagrangian (\ref{lagrangian}) describe the contribution of the proper inertia of the rotating string.
For the lowest approximation it is sufficient to retain only the
first terms (linear confinement) in the expansion of the string
terms in powers of $l_{1,2}^2$. Corrections to this potential regime coming from the further expansion of the string terms are a
footprint of the underlying string dynamics. The leading correction of this type, of order $l_{1,2}^2$, is known as the string correction
\cite{DKS} --- it will be taken into account later.

Then the zero--order Hamiltonian takes the form:
\be
H_0=\frac{\mu_q+\mu_{\bar{q}}+\mu_g}{2}+\frac{m^2+p_q^2}{2\mu_q}+
\frac{m^2+p_{\bar q}^2}{2\mu_{\bar q}}+
\frac{p_g^2}{2\mu_g}+\sigma |\vec {r}_q-\vec {r}_g|+
\sigma |\vec {r}_{\bar q}-\vec {r}_g|+V_{\rm Coul},
\label{h0}
\ee
where the long--range confining force is augmented by the short--range
Coulomb potential,
\be
V_{\rm Coul}=-\frac{3\alpha_s}{2|\vec {r}_q-\vec {r}_g|}-\frac{3\alpha_s}
{2|\vec {r}_{\bar q}-\vec {r}_g|}+\frac{\alpha_s}{6|\vec {r}_q-\vec
{r}_{\bar q}|}.
\label{coulomb}
\ee
The coefficients in Eq.~(\ref{coulomb}) correspond to the colour content of the
$q\bar{q} g$ system \cite{horn}.

The einbein fields are to be found from the constraint conditions \cite{einbein}:
\be
\frac{\partial H_0}{\partial \mu_q}=\frac{\partial
H_0}{\partial \mu_{\bar q}}=\frac{\partial H_0}{\partial
\mu_g}=0.
\label{extremum}
\ee
Thus, to quantise the system, one should find the einbeins from
Eq.~(\ref{extremum}) and substitute them back to the Hamiltonian (\ref{h0}). In such a
way, einbeins would become entangled functions of coordinates
and momenta. To avoid these complications, an approximate
einbein field method is used in the QCD string model calculations:
einbeins are treated as $c$--number variational parameters. The
eigenvalues of the spinless Hamiltonian (\ref{h0}) are found as functions
of $\mu_q$, $\mu_{\bar q}$, and $\mu_g$ and, finally, minimised with respect to the einbeins.
With such simplifying assumptions the spinless Hamiltonian $H_0$ takes an
apparently nonrelativistic form, with einbein fields playing the role of the
constituent masses of the quarks and the gluon. These quantities, however, are not
introduced as model parameters, but are calculated in a
relativistic formalism. Indeed, the procedure of taking extrema in the einbeins is nothing but the
summation of the entire series of relativistic corrections to the leading would-be nonrelativistic eigenenergies.

The einbein field method allows one to estimate corrections to the leading
potential regime (\ref{h0}). First, as was discussed before, one takes into account the
contribution of the string inertia, expanding the square roots in the
expression (\ref{lagrangian}) to the first order in $l_{1,2}^2$ --- this gives the string correction $V_{\rm str}$.
Second, employing the Feynman--Schwinger representation for the Green's functions
of spinning quarks and gluons one can extract the spin--dependent part of
interaction, as described in Ref.~\cite{sdterms}, hereafter denoted as $V_{\rm SD}$.
Finally, the FCM method allows one to calculate the nonperturbative selfenergy of the quarks which, as was shown in Ref.~\cite{selfen},
provides an overall shift of the hadron mass, as required by phenomenology. We use the notation $C$ for this contribution.
Thus the full form of the hybrid Hamiltonian reads:
\be
H=H_0+V_{\rm str}+V_{\rm SD}+C.
\label{H}
\ee

To specify extra terms in Eq.~(\ref{H}) let us first introduce various angular momentum operators:
\be
\vec{L}_{1g}=[\vec{r}_{31}\vec{p}_g],~\vec{L}_{2g}=-[\vec{r}_{23}\vec{p}_g],~
\vec{L}_{1q}=-[\vec{r}_{31}\vec{p_q}],~\vec{L}_{2\bar q}=[\vec{r}_{23}\vec{p}_{\bar q}],~
\vec{L}_q=[\vec{r}_{12}\vec{p}_q],~\vec{L}_{\bar q}=-[\vec{r}_{12}\vec{p}_{\bar q}],
\label{angularmomenta}
\ee
where
\be
\vec{r}_{12}=\vec{r}_q-\vec{r}_{\bar q},\quad\vec{r}_{31}=\vec{r}_g-\vec{r}_q,\quad\vec{r}_{23}=\vec{r}_{\bar q}-\vec{r}_g.
\label{r}
\ee

In terms of these coordinates and angular momenta the string correction $V_{\rm str}$ is trivially calculated from Eq.~(\ref{lagrangian})
and takes the form (due to the symmetry of the problem we set $\mu_q=\mu_{\bar q}=\mu$):
\be
V_{\rm str}=-\frac{\sigma}{6r_{13}}\left(\frac{L_{1g}^2}{\mu_g^2}
-\frac{\vec{L}_{1g}\vec{L}_{1q}}{\mu\mu_g}+\frac{L_{1q}^2}{\mu^2}\right)
+\genfrac{(}{)}{0pt}{0}{1\to 2}{q\to\bar{q}}.
\label{string}
\ee
The derivation of the spin--dependent potential can be found in Ref.~\cite{sdterms}, with the result:
\be
V_{\rm SD}=V_{\rm LS}^{(q\bar{q})}+V_{\rm LS}^{(g)}+V_{\rm SS}+V_{\rm ST}^{(q\bar{q})}+V_{\rm ST}^{(g)},
\label{sd}
\ee
where the superscript LS stands for the spin--orbit interaction, SS --- for the hyperfine interaction,
ST --- for the spin--tensor, and
\be
V_{\rm LS}^{(q\bar{q})}=-\frac{\alpha_s}{4\mu^2 r_{12}^5}\left(\vec{L}_q\vec{s}_q-\vec{L}_{\bar q}\vec{s}_{\bar q}\right),
\label{VLS0}
\ee
\be
V_{\rm LS}^{(g)}\hspace*{-0.5mm}=\frac{\sigma}{2r_{13}}\left(\frac{\vec{L}_{1q}\vec{s}_q}{\mu^2}-\frac{\vec{L}_{1g}\vec{s}_g}{\mu_g^2}
\right)+\frac{3\alpha_s}{2r_{13}^3}\left(\frac{1}{2\mu_g}+\frac{1}{\mu}\right)
\frac{\vec{L}_{1g}\vec{s}_g}{\mu_g}-\frac{3\alpha_s}{2r_{13}^3}\left(\frac{1}{2\mu}+\frac{1}{\mu_g}\right)
\frac{\vec{L}_{1q}\vec{s}_q}{\mu}+\genfrac{(}{)}{0pt}{0}{1\to 2}{q\to\bar{q}},
\label{so}
\ee
\be
V_{\rm SS}=-\frac{4\pi\alpha_s}{9\mu^2}\vec{s}_q\vec{s}_{\bar q}\delta(\vec {r}_{12})
+\frac{4\pi\alpha_s}{\mu\mu_g}\vec{s}_q\vec{s}_g\delta(\vec {r}_{13})
+\frac{4\pi\alpha_s}{\mu\mu_g}\vec{s}_{\bar q}\vec{s}_g\delta(\vec{r}_{23}),
\ee
\be
V_{\rm ST}^{(q\bar{q})}=-\frac{\alpha_s}{6\mu^2r_{12}^5}\left(3(\vec{s}_q\vec{r}_{12})
(\vec{s}_{\bar q}\vec{r}_{12})-r_{12}^2\vec{s}_q\vec{s}_{\bar q}\right),
\ee
\be
V_{\rm ST}^{\rm (g)}=\frac{3\alpha_s}{2\mu\mu_gr_{13}^5}\left(3(\vec{s}_q\vec{r_{13}})(\vec{s}_g\vec{r}_{13})-
r_{13}^2\vec{s}_q\vec{s}_g\right)+\genfrac{(}{)}{0pt}{0}{1\to 2}{q\to\bar{q}}.
\label{VST0}
\ee

Formally, these spin--dependent terms coincide with the well--known
Eichten--Feinberg--Gromes ones \cite{FEG}. Notice, however,
that the Eichten--Feinberg--Gromes spin-dependent potential comes out
from an expansion of
the interaction in the inverse powers of quark masses, whereas the potential $V_{\rm SD}$ derived above
contains effective constituent masses $\mu$ and $\mu_g$ in the denominators instead of current masses.
Due to confinement, these constituent-like masses
are always large, of order of the confinement scale ($\simeq\sqrt{\sigma}\simeq 400$ MeV) or larger,
even for massless particles.
So the result for $V_{\rm SD}$ is applicable to the case of light quark flavours, as well as to the
case of massless gluons. The only approximation made in order to derive the spin-dependent potentials
(\ref{VLS0})-(\ref{VST0}) is the Gaussian approximation for field correlators  --- see Ref.~\cite{Lisbon} for the discussion.

Finally, the constant $C$ is
\be
C=-\frac{3\sigma}{\mu\pi}\eta(m,T_g),
\label{constant}
\ee
where $\eta(m,T_g)$ is a universal function of the quark mass $m$ and the gluonic correlation length
$T_g$ --- see Ref.~\cite{selfen} for the explicit form of this function and for further details of the
formalism (in Ref.~\cite{selfen} the gluonic correlation length is denoted as $\delta$).

\section{The spectrum of hybrids}

The form (\ref{h0}) allows one to separate the centre-of-mass motion in a standard way, introducing the Jacobi coordinates as
\be
\vec{r}=\vec{r}_q-\vec{r}_{\bar{q}},\quad \vec{\rho}=\vec{r}_g-\frac{\mu_q\vec{r}_q+\mu_{\bar{q}}\vec{r}_{\bar{q}}}{\mu_q+\mu_{\bar q}}=
\vec{r}_g-\frac{\vec{r}_q+\vec{r}_{\bar{q}}}{2}.
\label{jacobi}
\ee

In terms of these Jacobi variables the Hamiltonian $H_0$ in the centre-of-mass frame can be written as
\be
H_0=\frac{m^2}{\mu}+\mu+\frac{\mu_g}{2}+\frac{p^2}{2\mu_{12}}+\frac{Q^2}
{2\mu_{12,3}}+\sigma r_{31}+\sigma r_{23}+V_{\rm Coul},
\label{cmh0}
\ee
where
\be
\mu_{12}=\frac{\mu}{2},\quad \mu_{12,3}=\frac{2\mu \mu_g}{M},\quad M=2\mu+\mu_g,
\label{mu}
\ee
and $\vec{p}$ and $\vec{Q}$ are the momenta conjugated to the Jacobi coordinates
$\vec{r}$ and $\vec{\rho}$, respectively.

First, we note that the relative angular momenta $l_{q\bar{q}}$ and $l_g$ are not conserved in the three--body
$q\bar{q}g$ system, though the requirement of a given $J^{PC}$ imposes restrictions on their possible values.
On the other hand, the zero--order Hamiltonian (\ref{cmh0}) conserves the total quark spin $s_{q\bar{q}}$. In
accordance with Eqs.~(\ref{JPCmag}) and (\ref{JPCel}), states with magnetic and electric gluons have different
total quark spin, so they are not mixed in the leading order. Thus one can employ the well--known hyperspherical
formalism to calculate the zero--order spectrum and w.f..

As we are interested in hybrids with a magnetic gluon, we use trial w.f.'s of the form
\be
|1^{--}\rangle_m=\Phi(r,\rho)S_0(q\bar{q})\sum_{\nu_1\nu_2}C^{1m}_{1\nu_1
1\nu_2}
\rho Y_{1\nu_1}(\hat {\rho})S_{1 \nu_2}(g),
\label{vector}
\ee
for the vector hybrid, and
\be
|J^{-+}\rangle_m=\Phi(r,\rho)\sum_{\mu_1\mu_2}C^{Jm}_{1\mu_1 1\mu_2}S_{1
\mu_1}
(q\bar{q})\sum_{\nu_1 \nu_2}C^{1 \mu_2}_{1 \nu_1 1 \nu_2}
\rho Y_{1 \nu_1}(\hat {\rho})S_{1 \nu_2}(g),
\label{siblings}
\ee
for its siblings. Here $S_{1\nu}(g)$ is the spin w.f. of the gluon, $S_0(q\bar{q})$ and
$S_{1\nu}(q\bar{q})$ are the singlet and triplet spin w.f.'s of the $q\bar{q}$ pair.
The ``radial" w.f. $\Phi(r,\rho)$ depends on its arguments in the form of the hyperspherical radius $R$,
\be
R^2=\frac{\mu_{12}}{M}r^2+\frac{\mu_{12,3}}{M}\rho^2.
\label{R}
\ee
Necessary formulae of hyperspherical formalism in three--body
systems can be found in Appendix~\ref{A} (see also Ref.~\cite{hyper} for more details).
In actual calculations the radial w.f. was chosen in the Gaussian form,
\be
\Phi(r,\rho)=\exp\left(-\frac12\beta^2MR^2\right),
\label{gauss}
\ee
with $\beta$ being the variational parameter. Then the eigenvalue $M_0$ is given by
\be
M_0=h(\mu_0,\mu_{g0},\beta_0),
\ee
where
\be
h(\mu,\mu_g,\beta)=\frac{\langle\Psi|H_0|\Psi\rangle}{\langle\Psi|\Psi\rangle},\quad
\Psi(\vec {r},\vec{\rho})=\rho Y_{1\nu}(\hat {\rho})\Phi(r,\rho).
\ee
and $\mu_0$, $\mu_{g0}$, and $\beta_0$ come as a selfconsistent solution of the set of three coupled equations,
\be
\frac{\partial h}{\partial\beta}=0,\quad\frac{\partial h}{\partial\mu}=0,\quad\frac{\partial h}{\partial \mu_g}=0.
\ee

The model parameters $\{m,\sigma,\alpha_s,\eta\}$ are fixed by evaluating the spectrum of conventional
$c\bar{c}$ $1S$-- and $1P$-- states within the same formalism (in Appendix~\ref{B} we give necessary details of calculations for the conventional
charmonium). The set of parameters is given in Table~\ref{t1}, and the results for charmonia levels are listed in Table~\ref{t2}.
The value of $\eta=0.29$ corresponds to the charmed quark mass given in Table~\ref{t1} and to the gluonic
correlation length $T_g\approx 0.2$ fm, which complies well with the one measured on the lattice (around $0.2\div 0.3$~fm \cite{tg}).

\begin{table}[t]
\begin{ruledtabular}
\begin{tabular}{ccccc}
Parameter&$m$, GeV& $\sigma$, GeV$^2$& $\alpha_s$&$\eta$\\
\hline
Value&1.48&0.16&0.55&0.29\\
\end{tabular}
\end{ruledtabular}
\caption{The set of parameters used for the numerical evaluation.}\label{t1}
\end{table}

\begin{table}[t]
\begin{ruledtabular}
\begin{tabular}{ccccccc}
\hline
State&${}^1S_0$&${}^3S_1$&${}^1P_1$&${}^3P_1$&${}^3P_0$&${}^3P_2$\\
\hline
Exp&2.980&3.097&3.526&3.511&3.415&3.556\\
\hline
Theor&2.981&3.104&3.528&3.514&3.449&3.552\\
\hline
\end{tabular}
\end{ruledtabular}
\caption{Masses of $S$- and $P$-level charmonia, in GeV.}\label{t2}
\end{table}

With these parameters the variational procedure described above yields
\be
M_0=4.573~{\rm GeV},
\label{M0}
\ee
for the zero--order hybrid mass, while the extremal values of the effective masses are
\be
\mu_0=1.598~{\rm GeV},\quad \mu_{g0}=1.085~{\rm GeV}.
\label{mu0}
\ee
Notice that the gluon effective mass $\mu_{g0}$ appears to be rather
large, and of the same order of magnitude as the effective charmed
quark mass $\mu_0$.

All corrections to the leading regime (\ref{M0}) are calculated as
perturbations, with the substitution $\mu \to \mu_0$, $\mu_g\to\mu_{g0}$, $\beta\to\beta_0$.
There are two types of such corrections: one that does not depend on quark
spin and the other, which depends. The former correction provides an overall shift
with respect to zero--order regime (\ref{M0}). The spin-dependent correction removes the
degeneracy between the four states (\ref{vector}) and (\ref{siblings}) and,
in principle, is responsible for the mixing of the magnetic gluon states
with electric gluon ones. In what follows we neglect such a mixing.

The simplest correction of the first type is the selfenergy
correction which shifts the zero--order hybrid mass downwards,
\be
\Delta M_{\rm selfenergy}=-28~{\rm MeV}.
\ee

The string correction does not depend on spins either, and it is calculated to be
\be
\Delta M_{\rm string}=-52~{\rm MeV}.
\ee

There is also a mass shift due to the gluon spin--orbit force,
common for all four states (\ref{vector}) and (\ref{siblings}), which
comes from the terms in (\ref{so}) proportional to the operator $\vec{s}_g$. This yields:
\be
\Delta M_{\rm LS}^{(g)}=-103~{\rm MeV}.
\ee

Finally, to calculate the spin splittings, it is convenient to rewrite the
operators $\vec{s}_q$ and $\vec{s}_{\bar q}$ in terms of operators
$\vec{s}_{q\bar{q}}$ and $\vec{\Sigma}$,
\be
\vec{s}_{q\bar{q}}=\frac{1}{2}(\vec{s}_q+\vec{s}_{\bar q}),\quad
\vec{\Sigma}=\frac{1}{2}(\vec{s}_q-\vec{s}_{\bar q}).
\ee
We notice then that, once the operator $\Sigma$ is antisymmetric with respect to the
permutation $q\leftrightarrow\bar{q}$, it flips the spin of the $q\bar{q}$ pair and,
as such, is neglected in our calculations. Then, after tedious but straightforward
calculations, one arrives at the following expressions for spin splittings:
\begin{eqnarray}
\Delta M(1^{--})&=&+\frac{3}{4}\Delta_{\rm SS}^{(qq)},\nonumber\\
\Delta M(0^{-+})&=&-\frac{1}{4}\Delta_{\rm SS}^{(qq)}-2\Delta,\nonumber\\
\Delta M(1^{-+})&=&-\frac{1}{4}\Delta_{\rm SS}^{(qq)}-\Delta,\\
\Delta M(2^{-+})&=&-\frac{1}{4}\Delta_{\rm SS}^{(qq)}+\Delta,\nonumber
\end{eqnarray}
where
\be
\Delta_{\rm SS}^{(qq)}=9~{\rm MeV}
\ee
comes from the quark--antiquark spin--spin interaction, and
\be
\Delta=\Delta_{SS}^{(qg)}+\Delta_{\rm LS}^{(q)}+\Delta_{\rm ST}.
\label{delta}
\ee
Individual contributions in Eq.~(\ref{delta}) are
\be
\Delta_{SS}^{(qg)}=9~{\rm MeV},\quad\Delta_{\rm LS}^{(q)}=24~{\rm MeV},\quad\Delta_{\rm ST}=35~{\rm MeV},
\ee
coming from the quark--gluon spin--spin interaction, the spin--orbit interaction
proportional to the quark and the antiquark spin, and the spin--tensor interaction, respectively. These all together give
\be
\Delta=68~{\rm MeV}.
\ee

The ultimate numerical results for the hybrid masses are given in
Table~\ref{t3}; spin splittings are established to yield:
\be
M(0^{-+})<M(1^{-+})<M(1^{--})<M(2^{-+}).
\label{pattern}
\ee

\begin{table}[t]
\begin{ruledtabular}
\begin{tabular}{ccccc}
$J^{PC}$&$0^{-+}$&$1^{-+}$&$1^{--}$&$2^{-+}$\\
\hline
Mass&4.252&4.320&4.397&4.457\\
\end{tabular}
\end{ruledtabular}
\caption{Masses of charmonium hybrids, in GeV.}\label{t3}
\end{table}

\section{Comparison to other approaches and lattice calculations}

The story of hybrid meson studies started with the bag model calculations
\cite{bag}, where the lowest charmonium hybrid mass of about 4~GeV was obtained,
and the splitting pattern (\ref{pattern}) was found.

In the flux tube model \cite{isgur} the hybrid excitations are visualised
as phonon--type excitations of the string connecting the quark--antiquark
pair, and a certain correspondence is established in Ref.~\cite{semay} between
the excited flux tube and the constituent gluon approaches.
Eight lowest $c\bar{c}g$ hybrids are predicted \cite{flux} to
reside around
$4.1\div 4.2$ GeV, and $1^{--}$ and $J^{-+}$ states are among those. However,
there is a discrepancy in quantum numbers of the flux tube hybrids and the
ones with a constituent gluon: the
constituent gluon carries colour and spin. As a result, the $P$--even flux tube hybrids have
\be
J^{PC}=0^{+-},1^{+-},2^{+-},1^{++}
\ee
quantum numbers, while in constituent gluon models these are
\be
J^{PC}=0^{++},1^{++},2^{++},1^{+-}
\ee
hybrids with the electric gluon. Spin splittings of the flux tube hybrids due to the long-range Thomas precession
were calculated in Ref.~\cite{merlin} to be small, while the splittings reported in the present paper
are much larger, and come mostly from perturbative short-ranged forces.

The constituent gluon model with pairwise forces was presented in the pioneering work
\cite{horn}, in Ref.~\cite{orsay}, and in Ref.~\cite{iddir}.
The mixing between magnetic and electric gluon hybrids was calculated
in Ref.~\cite{iddir}, with rather controversial results: in the first entry of Ref.~\cite{iddir} the
mixing was found to be small, while in the second entry it is claimed to be substantial (though
the details are not given there). The unmixed states are found in the same mass region as in
the present work.

There exist results \cite{buisseret} of the QCD string model
calculations in the einbein
field formalism. The $1^{-+}$ charmonium hybrid mass was found to be $4.2 \pm 0.2$ GeV.
However, this result cannot be directly compared with ours, as the
adiabatic approximation for quarks is employed there.

Finally, in a potential model, with the $c\bar{c}$ pair considered as a
colour--octet source, the tensor hybrid was predicted at 4.12 GeV
\cite{bicudo}.

Lattice simulations deal mostly with exotic quantum numbers, with $1^{-+}$ charmonium hybrid
residing at about $4.4$ GeV --- see Ref.~\cite{Michael} and references therein.
Among more recent results we would like to mention Ref.~\cite{liu1}, which gives
$4.405 \pm 0.038$ GeV for the $1^{-+}$ charmonium hybrid, and Ref.~\cite{liu2} where, for the first time,
a $1^{--}$ state was found, excited by the hybrid meson operator, with the mass of $4.379 \pm 0.149$ GeV
(the authors claim that they have found a radially excited vector hybrid; such an
interpretation was criticised in Ref.~\cite{burch}).
As seen from our Table~\ref{t3}, the agreement of lattice results and our findings is quite good.

\section{Hybrid decays}

The selection rule forbidding the decay of $1^{--}$ and $J^{-+}$ hybrids with magnetic gluon into the
$\bar D^{(*)}D^{(*)}$ final states was established for the constituent glue
model
in Refs.~\cite{orsay,myhybrdecay,kou}. As the decay takes place via {\em gluon $\to$
quark--antiquark pair} transition, the amplitude is proportional to the overlap of the
initial and final state w.f.'s with the pair creation operator. Thus it vanishes if
the final state mesons have identical spatial w.f.'s. The latter is definitely a good approximation
in the case of $\bar D^{(*)}D^{(*)}$.
Similar symmetry considerations hold true in the flux tube model \cite{hybrdecayflux}, and
seem to be rather general, as demonstrated in Ref.~\cite{page}.

There exists another selection rule based on the spin content of quarks in the final state,
established in Refs.~\cite{myhybrdecay,page} and, quite recently, in Ref.~\cite{burns}.
Assuming that i) the spin of the initial $q\bar{q}$ pair does not flip in
the decay, and ii) the $q\bar{q}$ pair created in the decay is in spin--triplet, one
can define the relative strength of matrix elements for decays into various final meson pairs.
This selection rule is rather powerful: for example, for a vector hybrid, the initial quark pair is
in spin--singlet, while for a conventional vector quarkonium it is in spin--triplet, with clear
discrimination between two possibilities. Thus, measuring
the relative rates of various $S$-- and $P$-- meson pairs, one could distinguish between
a vector quarkonium and a vector hybrid \cite{DK,burns}.

\begin{table}[t]
\begin{ruledtabular}
\begin{tabular}{ccccccccc}
&$\bar{D}D_0$&$\bar{D}^*D_0$&$\bar{D}D_1({}^1P_1)$&$\bar{D}^*D_1({}^1P_1)$
&$\bar{D}D_1({}^3P_1)$&$\bar{D}^*D_1({}^3P_1)$
&$\bar{D}D_2$&$\bar{D}^*D_2$\\
\hline
$1^{--}\vph$&&$\ds\frac{1}{\sqrt{6}}$&&$\ds-\frac12$&$\ds\frac12$&$\ds\frac{1}{2\sqrt{2}}$&
&$\ds-\frac12\sqrt{\frac56}$\\
\hline
$0^{-+}\vph$&$\ds-\frac{1}{\sqrt{2}}$&&&$\ds\frac{1}{\sqrt{2}}$&&&&\\
\hline
$1^{-+}\vph$&&$\ds-\frac{1}{\sqrt{3}}$&$\ds-\frac12$&$\ds\frac{1}{2\sqrt{2}}$&
$\ds\frac{1}{2\sqrt{2}}$&$\ds\frac14$&
&$\ds-\frac14\sqrt{\frac53}$\\
\hline
$2^{-+}\vph$&&&&$\ds-\frac{1}{2\sqrt{2}}$&&$\ds\frac34$&$\ds\frac{1}{2\sqrt{2}}$&$\ds\frac{\sqrt{3}}{4}$
\end{tabular}
\end{ruledtabular}
\caption{Spin--recoupling coefficients for the hybrid states listed in Table~\ref{t3}.
Here $D^{(*)}$ is an $S$--wave $D^{(*)}$-meson and $D_J$ is a $P$--wave
$D$--meson with the total momentum $J$.
A proper charge conjugation is implied.}\label{t4}
\end{table}

For hybrid decays under consideration,
only $S$--wave amplitudes are of immediate relevance (the $D$--wave ones are suppressed because
of a limited phase space), and the corresponding spin--recoupling coefficients are given in
Table~\ref{t4} (decay rates are proportional to the squares of these
coefficients).
The coefficients exhibit a sum rule: if the masses and w.f.'s
of the initial and final states involved could be taken identical, the total effect of coupling to open charm
mesons is identical within the whole hybrid multiplet (though individual contributions from various $D$--meson channels
differ). A similar sum rule holds true for the conventional $c\bar{c}$ charmonium couplings to $D$--mesons, as found
in Refs.~\cite{Eichten,Yusk} and formulated in Ref.~\cite{BS} as a general theorem. Most straightforward
consequence of this sum rule is the following observation: the effects of mesonic loops over the spectra could be
quite large, but numerically they are similar for all low--lying charmonia.
For hybrids, however, the latter is not the case as, in accordance with the results of Table~\ref{t3}, different members of hybrid multiplet reside among different thresholds.

\section{Discussion}

First, we notice that both lattice calculations and our findings place the vector hybrid at $4.4$ GeV,
substantially higher than the $Y(4260)$. One should have in mind, however, that the lattice result \cite{liu2}
for a vector hybrid comes with
a large error, and the accuracy of the einbein field method is not better than 5\% in the binding energy,
as estimated in Ref.~\cite{KNS}, so the discrepancy
could appear to be not very significant. On the other hand, the influence of open charm channels
is not taken into account in the present approach, and the said influence could be large.

Indeed, as seen from Table~\ref{t4}, there is a significant coupling of
the vector hybrid to the $\bar D D_1(^3P_1)$ channel.
There are two $D_1$-mesons, a narrow one with the mass of $2420$ MeV
and the width of $20$ MeV,
and a broad one with the mass of $2430$ MeV and the width of $380$ MeV
\cite{PDG},
which are (unknown) mixtures of the $^3P_1$ and $^1P_1$ states. Thus the vector hybrid state should be attracted to
the corresponding thresholds, which are tantalizingly close to the
measured mass of the $Y(4260)$.

It is interesting to mention in this regard another enigmatic vector state (or, maybe, even two states!), namely
$Y(4325)$ from BaBar \cite{4325} and $Y(4360)$ from Belle \cite{4360}, both seen in the initial state radiation process,
with the masses consistent with each other and with the width of the Belle state being two times smaller than that
of the BaBar one. The masses of these new $Y$'s are close to another
relevant threshold, the $\bar D^*D_0$ one (we assume,
following the Belle paper \cite{BelleD0}, the scalar $D$-meson to be at
$2308$ MeV, with the width of about $270$ MeV). So the coupling of the
hybrid vector
to $\bar D D_1$ and $\bar D^*D_0$ could be responsible for the formation
of two near--threshold states.

The case of the pseudoscalar hybrid is simpler: in accordance with
Table~\ref{t4}, half of the
decay strength goes to the $\bar DD_0$ channel with the nominal threshold
at $4.18$ GeV, so this hybrid would feed the
structure in the $\bar D D\pi$ final state, with this structure being
broad due to large $D_0$ width.

Exotic hybrid is estimated to be $80$ MeV lighter than the vector one,
with the mass very close to the $\bar D^*D_0$ threshold.
As seen from Table~\ref{t4}, the
coupling to $\bar D^*D_0$ is two times larger than the one for the vector
case, so,
as the $D_0(2308)$ is very broad, this hybrid has more chances to disappear
in the  $\bar D^*D\pi$ continuum.

As to the tensor hybrid, with the bare mass of $4.457$ GeV, it could survive as a resonance,
because the most prominent channel, $\bar D^*D_1$, opens nominally only
at $4.43$ GeV.

\section{Conclusions}

In this paper we calculated the masses of low--lying charmonium hybrids with magnetic gluon in the framework of
the Field Correlator Method for QCD. The QCD string approach is employed to estimate spin-dependent
corrections for the $1^{--}$, $0^{-+}$, $1^{-+}$, and $2^{-+}$ hybrid states.
The spectrum is calculated without fitting parameters, as
all the parameters of the effective Hamiltonian are fixed by reproducing $c\bar{c}$ charmonium
levels. Our results are in good agreement with lattice data.

Decay modes of hybrids are investigated and predictions are made for the relative rates of the decays into $S$- and $P$-wave
$D$-mesons for all four states of the lowest hybrid multiplet.

The calculated mass of the vector hybrid is $4.397$ GeV, substantially higher than the mass of
a promising hybrid candidate $Y(4260)$. We argue that strong coupling of the vector hybrid to the $DD_1$ and $D^*D_0$
modes can cause considerable threshold attraction, making vector hybrid bare state responsible for the
formation of near--threshold $Y(4260)$ and $Y(4325)$ states.

\begin{acknowledgments}
The authors acknowledge useful discussions with A. M. Badalian and Yu. A. Simonov.
This research was supported by the Federal Agency for Atomic Energy of
Russian Federation, by the grants RFFI-05-02-04012-NNIOa, DFG-436 RUS 113/820/0-1(R),
and by the Federal Programme of the Russian Ministry of Industry, Science, and
Technology No. 40.052.1.1.1112.
A. N. would also like to acknowledge the financial support via the project
PTDC/FIS/70843/2006-Fisica and of the non-profit ``Dynasty"
foundation and ICFPM.
\end{acknowledgments}

\appendix

\renewcommand{\theequation}{\thesection.\arabic{equation}}

\section{Some details of three--body kinematics}\label{A}

In a three--body system of particles with masses $\mu_1$, $\mu_2$,
$\mu_3$ three sets of Jacobi coordinates can be defined. One is
\be
\vec{r}_{12}=\vec{r}_1-\vec{r}_2,\quad\vec {\rho}_3=\vec{r}_3-\frac{\mu_1\vec {r}_1+\mu_2\vec {r}_2}{\mu_1+\mu_2},
\ee
and others ($\vec {r}_{31},\vec {Q}_2$ and $\vec {r}_{23},\vec {Q}_1$)
are obtained from it by cyclic permutations of the particle indices. The hyperspherical radius
\be
R^2=\frac{\mu_{12}}{M}r^2_{12}+\frac{\mu_{12,3}}{M}\rho_{3}^2,
\ee
where
\be
\mu_{12}=\frac{\mu_1\mu_2}{\mu_1+\mu_2},\quad \mu_{12,3}=\frac{(\mu_1+\mu_2)\mu_3}{M},
\ee
is invariant under such cyclic permutations. Similarly,
\be
d^3r_{12}d^3\rho_3=d^3r_{31}d^3\rho_2=d^3r_{23}d^3\rho_1.
\ee
Angular momenta (\ref{angularmomenta}) can be conveniently represented as
\begin{eqnarray}
\vec{L}_{1g}=\vec{r}_{31}\times\vec{p}_{31}-\frac{\mu_g}{\mu+\mu_g}\vec{r}_{31}\times\vec{Q}_{2},&\quad&
\vec{L}_{1q}=\vec{r}_{31}\times\vec{p}_{31}+\frac{\mu_g}{\mu+\mu_g}\vec{r}_{31}\times\vec{Q}_{2},\nonumber\\
\vec{L}_{2g}=\vec{r}_{23}\times\vec{p}_{23}+\frac{\mu_g}{\mu+\mu_g}\vec{r}_{23}\times\vec{Q}_{1},&\quad&
\vec{L}_{2q}=\vec{r}_{23}\times\vec{p}_{23}-\frac{\mu_g}{\mu+\mu_g}\vec{r}_{23}\times\vec{Q}_{1},\label{jacobiangmom}\\
\vec{L}_q=\vec{r}_{12}\times\vec{p}_{12}-\frac{1}{2}\vec{r}_{12}\times\vec{Q}_3,&\quad&
\vec{L}_{\bar q}=\vec{r}_{12}\times\vec{p}_{12}+\frac{1}{2}\vec{r}_{12}\times\vec{Q}_3,\nonumber
\end{eqnarray}
where the momenta $\vec {p}_{ij}$ and $\vec {Q}_k$ conjugated to corresponding Jacobi coordinates are introduced,
and $\mu_1=\mu_2=\mu$, $\mu_3=\mu_g$ is substituted.

The formula
\be
\vec {\rho}_3=\frac{M}{2(\mu+\mu_g)}\vec{r}_{31}-\frac{1}{2}\vec{\rho}_2=-\frac{M}{2(\mu+\mu_g)}\vec{r}_{23}-\frac{1}{2}\vec{\rho}_1
\ee
is extensively used in calculations of various matrix elements.

\section{Spectrum of conventional charmonia}\label{B}

In this Appendix we give some details of evaluation of the spectrum of conventional charmonia which was used in order to fix the set of
model parameters. Further details of calculations for conventional mesons can be found in Refs.~\cite{KNS,alla}.

The spinless centre-of-mass Hamiltonian of the charmonium reads:
\be
H^{(c\bar{c})}_0=2\sqrt{p^2+m^2}+\sigma r-\frac43\frac{\alpha_s}{r},
\label{Hcc}
\ee
where $r$ is the interquark distance. The einbein $\mu$ is introduced then,
\be
2\sqrt{p^2+m^2}\to\frac{m^2}{\mu}+\frac{\mu}{2}+\frac{p^2}{\mu}.
\ee

The variational procedure for the charmonium is similar to the one
described in the text body for the hybrid, that is, we take a Gaussian
trial w.f.,
\be
\Psi^{(c\bar{c})}(r)=\exp\left(-\frac12\mu\beta^2r^2\right),
\ee
and evaluate the average
\be
h^{(c\bar{c})}(\mu,\beta)=
\frac{\langle\Psi^{(c\bar{c})}|H_0^{(c\bar{c})}|\Psi^{(c\bar{c})}\rangle}{\langle\Psi^{(c\bar{c})}|\Psi^{(c\bar{c})}\rangle}.
\ee
Then, taking  extrema in the variational
parameters $\mu$ and $\beta$, we find the ``bare" charmonium mass
and the actual value $\mu_0$ to be substituted to the corrections to
the Hamiltonian (\ref{Hcc}). As in case of the hybrid, these corrections can be classified as spin-independent
and spin-dependent.

The spin-independent corrections to the Hamiltonian (\ref{Hcc}) are the string correction \cite{DKS},
\be
V_{\rm str}^{(c\bar{c})}=-\frac{\sigma L^2}{6\mu r},
\ee
with $\vec{L}$ being the quark--antiquark angular momentum, and the selfenergy correction $C$ which coincides with the one
for the hybrid, given by Eq.~(\ref{constant}) (this is due to the fact that the quark content is the same for both the charmonium and
charmonium hybrid \cite{selfen}).

The spin-dependent corrections are:
\be
V_{\rm SD}^{(c\bar{c})}=V_{\rm LS}^{(c\bar{c})}+V_{\rm SS}^{(c\bar{c})}+V_{\rm ST}^{(c\bar{c})},
\ee
where
\be
V_{\rm LS}^{(c\bar{c})}=-\frac{\sigma}{2\mu^2 r}\vec{L}\vec{S}+\frac{2\alpha_s}{\mu^2r^3}\vec{L}\vec{S},
\ee
\be
V_{\rm SS}^{(c\bar{c})}=\frac{32\pi\alpha_s}{9\mu^2}(\vec{s}_q\vec{s}_{\bar q})\delta(\vec{r}),
\ee
\be
V_{\rm ST}^{(c\bar{c})}=\frac{4\alpha_s}{3\mu^2r^3}
[3(\vec{s}_q\vec{r})(\vec{s}_{\bar q}\vec{r})-r^2(\vec{s}_q\vec{s}_{\bar q})],
\ee
with $\vec{s}_{q/\bar{q}}$ and $\vec{S}$ being the quark/antiquark spin and the total spin, respectively.

In Table~\ref{t2} we compare the predicted masses of the $1S$- and $1P$-wave charmonium states with the experimental data. The set of parameters used in the calculations is given in Table~\ref{t1}. We find a good agreement of our predictions with the data and use the same set of parameters to evaluate the masses of the hybrids.

\end{document}